\documentstyle[epsfig,preprint,aps]{revtex}
\begin{document}

\preprint{USM-TH-106}

\title{Photo-production of Nucleon
Resonances and Nucleon Spin Structure Function in the Resonance Region}
\author{Di Qing$^{1,2}$ and Iv\'an Schmidt$^1$}
\address{$^1$Departamento de F\'\i sica,
Universidad T\'{e}cnica Federico Santa Mar\'\i a,
Casilla 110-V, Valpara\'\i so, Chile}
\address{$^2$Institute of Modern Physics, Southwest Jiaotong University,
 Chengdu, 610031, China}
\maketitle

\begin{abstract}
The photo-production of nucleon resonances is calculated based on
a chiral constituent quark model including both relativistic
corrections $H_{\rm rel}$ and two-body exchange currents, and it
is shown that these effects play an important role. We also
calculate the first moment of the nucleon spin structure function
$g_1\left(x,Q^2\right)$ in the resonance region, and obtain a
sign-changing point around $Q^2\sim0.27~{\rm GeV}^2$ for the
proton.
\end{abstract}


\section{Introduction}

A major goal of hadron physics is to understand the spectrum, structure,
and interactions of hadrons as a consequence of the strong interactions.
However, quantum chromodynamics (QCD), the fundamental theory for strong
interactions, is so difficult to solve in the nonperturbative region of
low-energy hadron physics that no one is currently able to quantitatively
describe the hadron characteristics from first principles. Because of this,
effective models are built, trying to keep the essential features of the
underlying basic theory, and which hopefully can be solved or at least
approximated in a satisfactory way. Among these models, the constituent
quark model is the most successful. Nevertheless, it is important to
extend its range of application beyond the traditional static hadronic
properties, such as the mass spectrum, to processes that can provide
information in a more stringent way about the detailed quark dynamics.
Electromagnetic transitions play in this respect a special role, since
the photon is a particularly clean probe. On the experimental side, the
opening of CEBAF will certainly bring more precise experimental data in
the near future.

In its application to the photo-production of nucleon resonances,
the non-relativistic quark model had to be modified by adding some
terms in order to have a consistent theory. In fact, the study of
the photo-production of nucleon resonances has a long
history\cite{copley}. In the non-relativistic quark model, it is
begun with the work of Koniuk and Isgur,
based on one-body reduction current\cite{koniuk80}. Later on,
Close and Li stressed the importance of spin-orbit and
non-additive terms associated with the Wigner rotation of the
quark spin transformed from the frame of the recoiling quark to
the frame of the recoiling baryon, and also the role of the
binding potential\cite{close90} and configuration
mixing\cite{li90}. Since the spin-orbit (S-L) term of one-gluon
exchange potential\cite{rujula75} was ignored in the investigation
of baryon spectrum in the Isgur model\cite{isgur78}, most of the
authors did not consider the two-body exchange currents
corresponding to the one-gluon exchange potential in the
calculation of resonance photon decay. However, the S-L term is
important in the study of baryon-baryon interactions. Some authors
also included it in order to investigate the baryon
spectroscopy\cite{mukherjee93}.

On the other hand, chiral symmetry is probably the most important
feature of QCD in the nonperturbative region\cite{manohar84}. This
is introduced in the quark model by adding one-pion and one-sigma
exchange potentials between constituent quarks. Recently, based on
this chiral constituent quark model, Buchmann {\it et al.}
investigated the electromagnetic properties of octet and decuplet
baryons, including the two-body gluon and meson exchange currents.
In their calculations, the two-body exchange currents play an
important role in the explanation of the nonzero $\gamma N
\rightarrow \Delta$ $E2$ transition\cite{buchmann97} and the
neutron charge radius\cite{buchmann91}, and also in order to fit
the electromagnetic properties of the octet and decuplet
baryons\cite{wagner95}. Meyer {\it et al.} included the two-body
exchange currents to study the photo-production of the resonance
states below $\sqrt{s}=1.6$ GeV and obtained a reasonable good
agreement with experiment\cite{meyer97}.

In the first part of this work, based on the chiral constituent
quark model\cite{buchmann91}, we go beyond the work of Meyer {\it
et al.}\cite{meyer97}, calculating the photo-production amplitudes
of nucleon resonances up to energies higher than they do, that is
up to $2~{\rm GeV}$, and we also include the contributions of the
spin-orbit and non-additive terms, in addition to the
contributions of the two-body exchange currents.

The effect of nucleon resonances is also important in other
electromagnetic transition processes. In fact, recently there has
been much interest in the nucleon spin structure function
$g_1\left(x,Q^2\right)$ in the low energy region, where it is
mainly saturated by the low-lying resonances. For the proton,
since $g_1\left(x,Q^2\right)$ satisfies a positive sum rule in the
large $Q^2$ limit, and the negative Gerasimov-Drell-Hearn (GDH)
sum rule in the real photon limit\cite{gdh}, there should be a
sign-changing in the low $Q^2$ region. Theoretically,
phenomenological models\cite{burkert94,li96,ma98,drechsel99} have
been proposed to study the $Q^2$ evolution of the first moments of
the nucleon spin structure function in the resonance region. After
taking into account the nucleon resonances explicitly, the
predictions\cite{burkert94} are in good agreement with the few
data points of the SLAC experiment E143, although within large
experimental errors\cite{E14397}, and gave a sign-changing point
at a smaller $Q^2\sim 0.22~{\rm GeV}^2$, instead of $0.8~{\rm
GeV}^2$\cite{burkert93} that was predicted before. Li and Dong
considered the interference cross section between transverse and
longitudinal currents based on a non-relativistic constituent
quark model. In their calculation, the contribution of the
interference cross section pushes the sign-changing point to
around $0.53~{\rm GeV}^2$\cite{li96}. After including the
background contribution, the sign-changing point appears at a
smaller energy scale, about $0.3~{\rm GeV}^2$\cite{ma98}. Soffer
and Teryaev introduced a parametrization to investigate the $Q^2$
dependence of GDH sum rule\cite{soffer93}. All these calculations
indicate that there is a very strong $Q^2$ dependence behavior on
the evolution of the nucleon spin structure function due to the
resonance contributions. In order to reliably perform the
saturation, it is necessary to take into account all kinds of
effects carefully, and that is what we plan to do here.

In section II of this paper, we calculate the photo-production amplitudes
of nucleon resonances up to energies of $2~{\rm GeV}$, including both the
relativistic corrections and the two-body exchange currents. Furthermore,
we also study the effects of the two-body exchange currents in the nucleon
spin structure function and GDH sum rule in the resonance region in
section III. Finally a summary and conclusions are given in section IV.

\section{The photo-production of nucleon resonances}

The helicity amplitudes of nucleon excitations are defined as
\begin{equation}
A_\lambda = \langle R, J^\prime, \lambda \left | H_T \right | N, \frac{1}{2},
\lambda - 1 \rangle,
\end{equation}
where $\lambda$ is the helicity of resonance $R$ with total angular
moment $J^\prime$. The electromagnetic transition operator $H_T$ is
composed of one-body non-relativistic reduction term, spin-orbit and
non-additive terms (the last two terms, first introduced by Brodsky
and Primack\cite{brodsky69}, are called collectively as
$H_{\rm rel}$)\cite{close90},
\begin{equation}
H_T = H_{NR} + H_{SO} + H_{NA}.
\end{equation}
In fact, it is more convenient to calculate the
transition amplitudes in the form of current matrix elements. For real
photons, we have
\begin{equation}
A_\lambda = -\sqrt{\frac{2\pi}{\omega}}\langle R, J^\prime, \lambda \left |
\vec{\epsilon}\cdot\vec{j}\left( \vec{q} \right)
\right | N, \frac{1}{2}, \lambda -1 \rangle,
\end{equation}
in which $\vec{\epsilon}=-\frac{1}{\sqrt{2}}\left ( 1,i,0 \right )$
is the photon transverse polarization vector,
$\left(\omega,\vec{q}\right)$ is the four momentum transfer
of the photon in the center of mass frame,
$\vec{j}\left( \vec{q} \right)$ is the electromagnetic current
corresponding to the transition operator $H_T$, and which can be
easily obtained from $H_T$ as:
\begin{eqnarray}\label{close}
\vec{j}\left( \vec{q} \right) &=& \vec{j}_{NR}\left( \vec{q} \right)
+ \vec{j}_{OS}\left( \vec{q} \right) + \vec{j}_{NA}\left( \vec{q} \right), \\
\vec{j}_{NR}\left( \vec{q} \right) &=& \sum^3_{i = 1}\mu_i
\left ( i\left [ \vec{\sigma}_i
\times \vec{p}_i, e^{i\vec{q}\cdot\vec{r}_i}\right ] +
\left \{ \vec{p}_i, e^{i\vec{q}\cdot\vec{r}_i}\right \} \right ),\label{enr} \\
\vec{j}_{SO}\left( \vec{q} \right) &=& \sum^3_{i = 1}-\frac{1}{2}
\left [ 2\mu_i - \frac{e_i}{2 m}
\right ] \frac{\omega}{2 m_i}i\left \{ \vec{\sigma}_i\times\vec{p}_i,
e^{i\vec{q}\cdot\vec{r}_i}\right \}, \\
\vec{j}_{NA}\left( \vec{q} \right) &=& \sum^3_{i<j} \frac{\omega}{4 M_T}
i \left ( \frac{\vec{\sigma}_i}{m}- \frac{\vec{\sigma}_j}{m} \right )
\times\left( e_j e^{i\vec{q}\cdot\vec{r}_j}\vec{p}_i - e_i
e^{i\vec{q}\cdot\vec{r}_i}\vec{p}_j \right ).
\end{eqnarray}
Here, $e_i$, $\vec{p}_i$, $\vec{r}_i$, $\mu_i$ are the charge, momentum,
position and magnetic moment of the {\it i}th quark, $m$ and $M_T$ are
the quark and nucleon mass, and $A_i$, $\phi_i$, $B_i$ and $E_i$ are the
electromagnetic fields. According to Eq.(\ref{enr}),
we can see that $\vec{j}_{NR}$ is exactly the lowest order non-relativistic
one-body electromagnetic current reduction. With the multipole expansion
of electromagnetic current\cite{deforest66}
\begin{eqnarray}\label{multi}
j\left(\vec{q};m\right) = \left \{
\begin{array}{ll}
\frac{\omega}{\left| \vec{q}\right |}\sqrt{4 \pi}\sum_{J\ge 0}i^J
\sqrt{2 J + 1} \hat{M}_{Jm}\left ( \left| \vec{q}\right |\right)
& \,\,\,\,    m = 0, \\
-\sqrt{2\pi}\sum_{J\ge 1}i^J \sqrt{2 J + 1}\left(
\hat{T}^{\rm el}_{Jm}\left ( \left| \vec{q}\right |\right) +
m\hat{T}^{\rm mag}_{Jm}\left ( \left| \vec{q}\right |\right)\right)
& \,\,\,\, m=\pm 1,
\end{array} \right.
\end{eqnarray}
in which $\hat{M}_J$, $\hat{T}^{\rm el}_J$ and $\hat{T}^{\rm mag}_J$ are
Coulomb, electric and magnetic multipole operators of rank $J$,
respectively, we have
\begin{eqnarray}
A_\lambda =&& \frac{2 \pi}{\sqrt{\omega}}\sum_{J \ge 1}i^J\sqrt{2J+1}
\left(\langle R,J^\prime, \lambda \left|
\hat{T}^{\rm el}_{J1}\left ( \left| \vec{q}\right |\right)
\right |N, \frac{1}{2}, \lambda-1\rangle \right. \nonumber \\
&& \left. +\langle R,J^\prime, \lambda \left|
\hat{T}^{\rm mag}_{J1}\left ( \left| \vec{q}\right |\right)
\right |N, \frac{1}{2}, \lambda-1\rangle \right ).
\end{eqnarray}
Due to the parity and angular momentum conservation, only one
electric and one magnetic multipole operators contribute to the
helicity amplitudes.

In this work we study the photo-production amplitudes of nucleon
resonances based on the chiral constituent quark
model\cite{buchmann91}, which includes not only the one-gluon
exchange potential, but also the one-pion and one-sigma exchange
potentials to model chiral symmetry breaking of QCD in the
nonperturbative region. As pointed out in the introduction, the
S-L term in the one-gluon exchange potential is important in
baryon-baryon interactions. Furthermore, when it is included in
the study of the baryon spectrum, and because of gauge invariance,
it is necessary to consider its effects, namely the two-body gluon
exchange current $\vec {j}_{gq{\bar q}}\left(\vec{r}_i,\vec{r}_j,
\vec{q}\right)$, which turns out to be important in the
investigation of the electromagnetic transitions\cite{buchmann91}.
There exist several methods to derive the two-body gluon exchange
current, such as the minimal electromagnetic coupling method and
the Feynman diagram method. Buchmann {\it et al.} derived it by
the Feynman diagram approach\cite{buchmann91}. In addition, the
gauge invariant currents corresponding to the one-pion exchange
potential are the pion-pair current $\vec{j}_{\pi q {\bar q}}
\left(\vec{r}_i,\vec{r}_j,\vec{q}\right)$ and the pionic current
$\vec{j}_{\gamma \pi\pi}\left(\vec{r}_i,\vec{r}_j,\vec{q}\right)$,
which are also derived using Feynman diagrams. Explicit
expressions for all these two-body exchange currents can be found
in Ref. \cite{buchmann91}.

In Ref. \cite{close90}, the quark mass was treated as free parameter
in order to include the contributions of the scalar potentials. Here,
we will investigate the contributions of the scalar potentials carefully.
Buchmann {\it et al.} found that the scalar potentials contribute
significantly to the electromagnetic properties of the octet and decuplet
baryons. According to Eq.(2.10) of Ref. \cite{close90}, to
$O\left(1/m^2\right)$ we have
\begin{equation}
\vec{j}_{\rm scalar}\left(\vec{r}_i,\vec{r}_j,\vec{q}\right)
= -\frac{i}{2m^2}\left\{e_i e^{i\vec{q}\cdot\vec{r}_i}
\vec{\sigma}_i\times\vec{q}V^{\rm scalar}\left(\vec{r}_i,
\vec{r}_j\right) + \left(i\leftrightarrow j\right) \right\}.
\end{equation}
In the chiral constituent quark model\cite{buchmann91}, the scalar potentials
include a two-body harmonic oscillator confinement potential
$V^{conf}\left(\vec{r}_i,\vec{r}_j\right)$ and a one-sigma exchange
potential $V^{\sigma}\left(\vec{r}_i,\vec{r}_j\right)$.

Thus, in the chiral constituent quark model\cite{buchmann91}, the total
electromagnetic current $\vec{j}_{tot}$, which satisfies the gauge
invariance condition, should be given by a sum of one-body non-relativistic
reduction, spin-orbit, non-additive, and two-body exchange currents:
\begin{eqnarray}
\vec{j}_{tot} &=& \vec{j}_{NR}\left( \vec{q} \right) + \vec{j}_{OS}
\left( \vec{q} \right) + \vec{j}_{NA}\left( \vec{q} \right) +
\sum_{i<j=1}^{3}\left(\vec{j}_{gq{\bar q}}\left(\vec{r}_i,\vec{r}_j,
\vec{q}\right) +
\vec{j}_{\pi q {\bar q}}\left(\vec{r}_i,\vec{r}_j,\vec{q}\right)\right.\nonumber \\
&&\left. +
\vec{j}_{\gamma \pi\pi}\left(\vec{r}_i,\vec{r}_j,\vec{q}\right) +
\vec{j}_{conf}\left(\vec{r}_i,\vec{r}_j,\vec{q}\right) +
\vec{j}_{\sigma}\left(\vec{r}_i,\vec{r}_j,\vec{q}\right)\right).
\end{eqnarray}

In our calculations the model wave function is the standard
$SU\left(6\right)\otimes O\left(3\right)$, all the parameters
are the same as in Ref.\cite{buchmann91}, and the phase convention
of Koniuk and Isgur\cite{koniuk80,close90} is used. Our main purpose
is to investigate the role of the two-body exchange currents, and
spin-orbit and non-additive terms, in the photo-production of
nucleon resonances. As a first step, we will neglect the
configuration mixing effects. We also neglect the finite electromagnetic
size of constituent quarks used in Ref.\cite{buchmann97,meyer97}.
In fact, this is easy to take into account, by multiplying every
photo-production amplitude by a monopole form factor
$F_{\gamma q}\left(\vec{q}^2\right)= 1/\left(1+\left(1/6\right)
\vec{q}^2 r^2_{\gamma q}\right)$.

All the calculations are made in the center mass frame and the
results are shown in the Tables \ref{tab1} and Tables \ref{tab2}.
The experimental data are given by the most recent Particle Data
Review of Particle Physics\cite{pdg}. We can see that the two-body
exchange currents
play an important role, especially the confinement current, in the
photo-production of nucleon resonances. For the resonance
$S_{11}\left(1535\right)$, in contrast to Ref. \cite{close90,li90},
the theoretical overestimate versus experimental result does not
exist anymore. It is interesting to notice that the Roper resonance
$P_{11}\left(1440\right)$ agrees with experiment due to the two-body
exchange currents, and not to the configuration mixing\cite{li90}.
Another example is the $D_{15}\left(1675\right)$ proton excitation
amplitudes $A_{1/2}$ and $A_{3/2}$, for which the two-body exchange
currents give a nonzero contribution, although it is not in the
right direction. However, for $P_{33}\left(1232\right)$, the
theoretical result is still far below the experimental data because
the two-body exchange currents cancel each other. Comparing with
the large contributions to the $L=0$ and $L=1$ resonances, the
contributions of the two-body exchange currents to the $L=2$
resonances are smaller, but they are still important to some
$L=2$ resonances.

On the other side, we confirm the results of Close and Li\cite{close90},
that the relativistic corrections $H_{\rm rel}$ are important in most
cases. Meyer {\it et al.} stressed the role of the two-body exchange
currents on the photo-production of the resonance states with
masses below $\sqrt{s}=1.6~{\rm GeV}$\cite{meyer97}, but the relativistic
corrections $H_{\rm rel}$ are neglected in their calculations.
In fact, after considering the contributions of both the two-body
exchange currents and $H_{\rm rel}$, the agreement with experimental
is improved, specially for the resonances
$D_{13}\left(1520\right)$ and $P_{33}\left(1232\right)$.

\section{The nucleon spin structure in the resonance region}

The nucleon spin structure in the resonances region is mainly saturated
by the nucleon resonances of photo- and electro-production. Extending the
calculation of the photo-production to electro-production is straightforward.
However, the electro-production of resonances will become highly
relativistic when the momentum transfer $q^2$ increases.
We will follow the method of Li and Close\cite{li90} in order to remove the
exponential decay with $q^2$. This method was first used by Foster
and Hughes in order to study the electro-production in the equal-velocity
frame\cite{foster82}.

In general, the nucleon spin structure function $g_1\left(x, Q^2\right)$
can be expressed as\cite{ioffe83}
\begin{equation}
g_1\left(\omega, Q^2\right)=\frac{M_T K}{8\pi^2 \alpha
\left(1+\frac{Q^2}{\omega^2}\right)}\left[ \sigma_{1/2}
\left(\omega,Q^2\right)-\sigma_{3/2}\left(\omega,Q^2\right)+
\frac{2\sqrt{Q^2}}{\omega}\sigma_{TL}\left(\omega,Q^2\right)
\right].
\end{equation}
Here, $\sigma_{3/2}$ and $\sigma_{1/2}$ are the total cross sections,
related to the excitation of hadronic states with helicity $3/2$ and
$1/2$ respectively, $\sigma_{TL}$ is the longitudinal and transverse
interference cross section, and $x=Q^2/2M_T\omega$ is the Bjorken
scaling variable. Following Ref. \cite{drechsel99n}, we choose the
photon flux $K$ as ``photon equivalent energy''
 $K=\left(W^2-M_T^2\right)/2M_T$, where $W$ is the total c.m. energy.
Then we have for the first moment of the nucleon spin structure function
\begin{eqnarray}
\Gamma\left(Q^2\right)&=&\int^{x_0}_0 dx\, g_1\left(x,Q^2\right)\nonumber\\
&=&\frac{Q^2}{16\pi^2 \alpha}\int^{\infty}_{\omega_{th}}\frac{1-x}
{\sqrt{\omega^2+Q^2}}\left[ \sigma_{1/2}
\left(\omega,Q^2\right)-\sigma_{3/2}\left(\omega,Q^2\right)+
\frac{2\sqrt{Q^2}}{\omega}\sigma_{TL}\left(\omega,Q^2\right)
\right]\frac{d\omega}{\omega},
\end{eqnarray}
where $\omega_{th}$ is the threshold energy of pion photo-production,
and $x_0=Q^2/2M_T\omega_{th}$. In the large $Q^2$ limit,
$\Gamma\left(Q^2\right)\rightarrow \Gamma$ (constant), and it
is still positive at $Q^2=0.5~{\rm GeV}^2$\cite{E14397}.
However, in the real photon limit, the cross section $\sigma_{TL}$ vanishes,
and in fact the negative GDH sum rule\cite{gdh} must be satisfied
\begin{equation}
I\left(Q^2=0\right)=\int_{\omega_{th}}^\infty
\frac{d\omega}{\omega} \left[\sigma_{1/2}\left(\omega,Q^2=0\right)
-\sigma_{3/2}\left(\omega,Q^2=0\right)\right] =
-\frac{2\pi^2\alpha}{M_T^2} \kappa^2,
\end{equation}
where $\kappa$ is the anomalous magnetic moment of the nucleon,
and $I\left(Q^2\right)=\left(2M_T^2/Q^2\right)\Gamma\left(Q^2\right)$.
Thus the nucleon spin structure is strongly $Q^2$ depended in the
resonance region.

With a Breit-Wigner parametrization in the resonance region,
the absorption cross section can be expressed in terms of amplitudes
as\cite{stoler93}
\begin{eqnarray}
\sigma_{1/2}\left(\omega,Q^2\right) &=& \sum_R\frac{2M_T}{\left(W+W_R\right)}
\frac{\Gamma_R}{\left(W-W_R\right)^2 + \Gamma_R^2/4}
\left| A_{1/2}^R\left(\omega,Q^2\right)\right|^2, \\
\sigma_{3/2}\left(\omega,Q^2\right) &=& \sum_R\frac{2M_T}{\left(W+W_R\right)}
\frac{\Gamma_R}{\left(W-W_R\right)^2 + \Gamma_R^2/4}
\left| A_{3/2}^R\left(\omega,Q^2\right)\right|^2, \\
\sigma_{TL}\left(\omega,Q^2\right) &=& \sum_R\frac{2M_T}{\left(W+W_R\right)}
\frac{\Gamma_R}{\left(W-W_R\right)^2 + \Gamma_R^2/4}
\left[A_{1/2}^{R*} S_{1/2}^R +  S_{1/2}^{R*} A_{1/2}^R\right]
\end{eqnarray}
where, $\Gamma_R$ is the total decay width of the resonance $R$.
The longitudinal transition amplitude is
\begin{equation}
 S_{1/2} = \langle R, J^\prime, \frac{1}{2} \left | H_l \right | N,
\frac{1}{2},\frac{1}{2} \rangle,
\end{equation}
with the longitudinal transition operator
\begin{equation}\label{eqhl}
H_l = \frac{1}{\sqrt{2\omega}}\left(\epsilon_0 j_0 - \epsilon_3 j_3\right),
\end{equation}
which is slightly different from Eq.(11) of Ref. \cite{li96} with
a factor $1/\sqrt{2\omega}$,
and which will lead to a more standard form of charge operators
in the following. The longitudinal polarization vector can be
chosen as
\begin{equation}\label{vector}
\epsilon_\mu = \left(\epsilon_0,0,\epsilon_3\right)
=\left(\frac{q_3}{\sqrt{Q^2}},0,0,\frac{\omega}{\sqrt{Q^2}}\right).
\end{equation}
Combining Eq.(\ref{eqhl}) with Eq. (\ref{vector}), we have
\begin{equation}
 S_{1/2} = \frac{\sqrt{Q^2}}{\omega\sqrt{2\omega}}\langle R, J^\prime,
\frac{1}{2} \left | j_3 \right | N, \frac{1}{2}, \frac{1}{2} \rangle.
\end{equation}
Substituting Eq.(\ref{multi}) into the above equation, we obtain
\begin{equation}
 S_{1/2} = \frac{\sqrt{Q^2}}{\left|\vec{q}\right|\sqrt{2\omega}}
\sqrt{4 \pi}\sum_{J\ge 0}i^J\sqrt{2 J + 1}\langle R, J^\prime,
\frac{1}{2} \left |\hat{M}_{Jm}\left ( \left| \vec{q}\right |\right)
\right | N, \frac{1}{2}, \frac{1}{2} \rangle.
\end{equation}
Also, according to the parity and angular momentum conservation,
only one Coulomb multipole operator contributes to the
longitudinal transition amplitude.

The charge operator, corresponding to the current in Eq.(\ref{close}),
is\cite{close92}
\begin{eqnarray}
\rho\left(\vec{r}_i,\vec{r}_j,\vec{q}\right) &=&
 \sum_i \left[e_i + i \frac{e_i}{4m^2}\vec{q}\cdot \left(\vec{\sigma}_i
\times\vec{p}_i\right)\right] e^{i\vec{q}\cdot\vec{r}_i}
  \nonumber \\
 &-&
 \sum_{i<j}\frac{i}{4M_T}
\left( \frac{\vec{\sigma}_i}{m}-\frac{\vec{\sigma}_j}{m}\right)
\left[e_i\left(\vec{q}\times\vec{p}_j\right) e^{i\vec{q}\cdot\vec{r}_i}
-e_j\left(\vec{q}\times\vec{p}_i\right) e^{i\vec{q}\cdot\vec{r}_j}\right],
\end{eqnarray}
which includes both the spin-orbit and no-additive terms that are known
to be important in reproducing the sum rule for the quantity
$\sigma_{TL}$\cite{li96}. The factor $1/{2\omega}$, appeared in the charge
operator of Ref. \cite{li96,close92}, is absorbed into Eq. (\ref{eqhl}),
and the first term in the above equation is just the usual charge operator.
All of those charge operators related to the two-body exchange currents
are $O\left(1/m^3\right)$, and can be
neglected in the calculation of $\sigma_{TL}$, although they are
crucial in the case of the $\gamma N\rightarrow\Delta$ $E2$
transition\cite{buchmann97} process and in reproducing the
neutron charge radius\cite{buchmann91}.

In our calculations all the nucleon resonances with masses below
$2.0~{\rm GeV}$ are included. Furthermore, in order to connect
smoothly with the GDH sum rule $I\left(Q^2\right)$ at $Q^2 = 0$
and  with the sum rules for $\Gamma_{p,n}\left(Q^2\right)$ at high
$Q^2$, and based on the assumption that the main contribution
comes from vector meson intermediate states at $Q^2 \le 2 - 3 {\rm
GeV}^2$, the non-resonant part is modeled as in Ref.
\cite{burkert94}:
\begin{eqnarray}
I^{nres}(Q^2) &=& 2m^2 \Gamma\left(Q^2_m\right)\left[ \frac{1}{Q^2
+ \mu^2} - \frac{c\mu^2}{\left(Q^2 + \mu^2\right)^2}\right], \\ c
&=& 1 +
\frac{\mu^2}{2m^2}~\frac{1}{\Gamma\left(Q^2_m\right)}\left[
\frac{1}{4}\kappa^2 + I^{res}\left(0\right)\right],
\end{eqnarray}
where $\Gamma\left(Q^2_m\right)$ is a free parameter, $c$ is fixed
by the value of the GDH sum rule at $Q^2=0$, and $\mu$ is the mass
of $\rho$ meson. The extended GDH sum rule is simply expressed as
a sum of the resonant part $I^{res}\left(Q^2\right)$ and the
non-resonant part $I^{nres}\left(Q^2\right)$
\begin{equation}
I\left(Q^2\right)=I^{res}\left(Q^2\right) +
I^{nres}\left(Q^2\right).
\end{equation}
In fact, the importance of the non-resonant part was stressed
early by Karliner\cite{karliner73}, in order to reproduce the GDH
sum rule at $Q^2 = 0~{\rm GeV}^2$. In our calculations, the chosen
parameters are $\Gamma_p\left(Q^2_m\right)=0.237$ and
$\Gamma_n(Q^2_m) = -0.025$, and $c_p = 1.387$ and $c_n = -2.753$.
These numbers were obtained using the experimental results for
$\Gamma_{p,n}\left(Q^2\right)$ at $Q^2 = 2~{\rm GeV}^2$, and by
the values of the GDH sum rules $I_{p,n}\left(Q^2=0\right)$,
respectively. They are somewhat different from those of Ref.
\cite{burkert94}, but the changing in the shape of
$\Gamma_{p,n}(Q^2)$ is weak in the low $Q^2$ region.

\begin{figure}
\begin{center}
\epsfig{figure=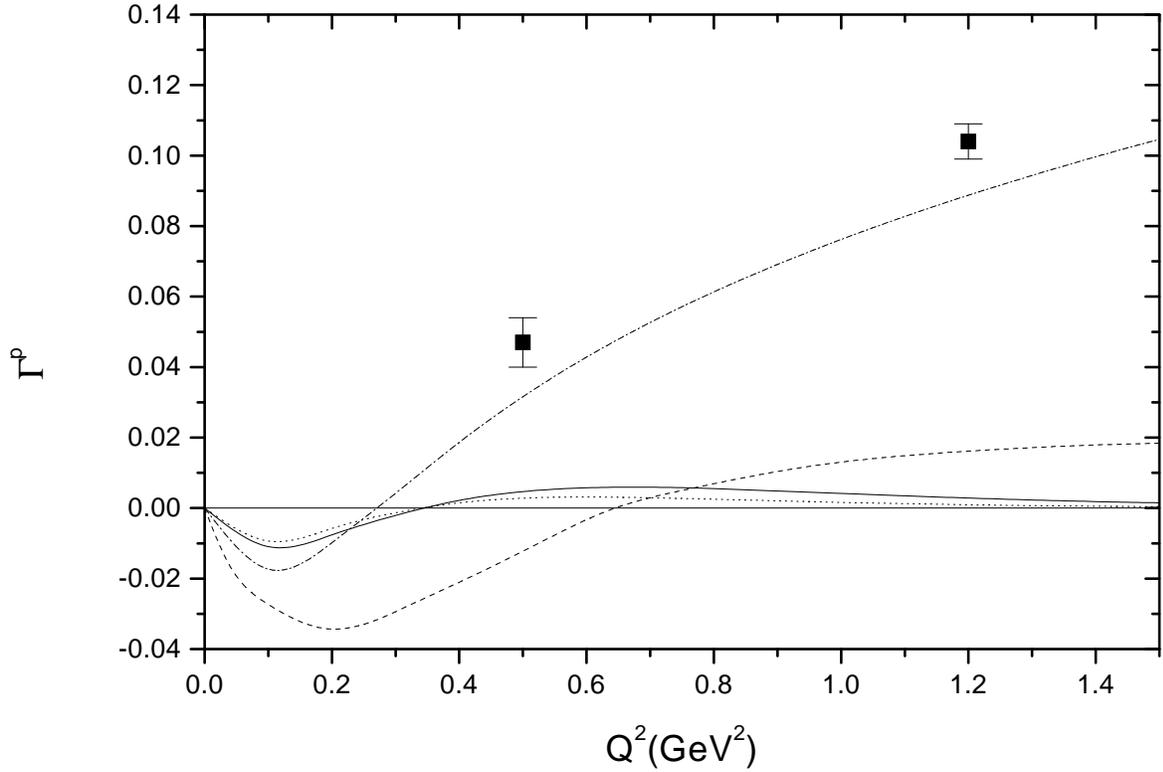, height=13cm} \caption{The first moment of
the spin structure function $g_1\left(x, Q^2\right)$ for proton.
The solid and dotted lines show our calculation with and without
the two-body exchange currents. The dashed line is the result of
Ref. \protect\cite{burkert94}. The dashed-dotted line is our
result with the non-resonant part and the two-body exchange
currents contributions. The experimental data are from
\protect\cite{E14398}.}\label{g1p}
\end{center}
\end{figure}

\begin{figure}
\begin{center}
\epsfig{figure=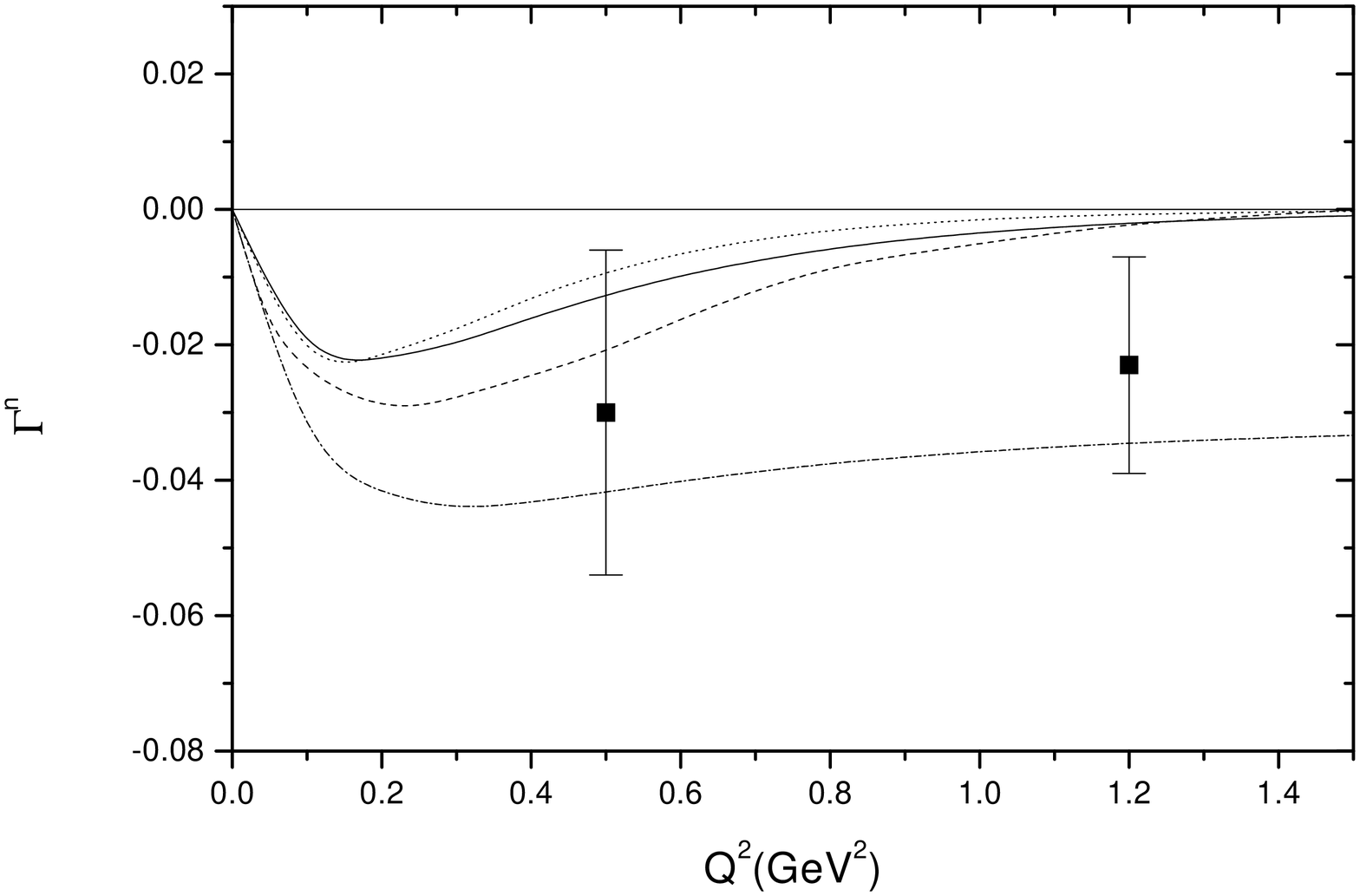, height=13cm}
\caption{The same as Fig. \ref{g1p} for neutron.}\label{g1n}
\end{center}
\end{figure}

The $Q^2$ dependence of the first moment of the nucleon spin
structure function $g_1\left(x, Q^2\right)$, both for the proton
and the neutron in the resonance region, is shown in Figs.
\ref{g1p} and \ref{g1n}. We include the resonant part calculation
of Burkert and Ioffe\cite{burkert94}, for comparison. The main
difference between this work and ours is that we use the chiral
constituent quark model, which in practice implies a different set
of parameters, and we also include two-body exchange currents,
although in contrast to the photo-production of nucleon
resonances, in the case of the first moment of the nucleon spin
structure function $g_1\left(x, Q^2\right)$ the effect of the
two-body exchange currents is small.

\begin{figure}
\begin{center}
\epsfig{figure=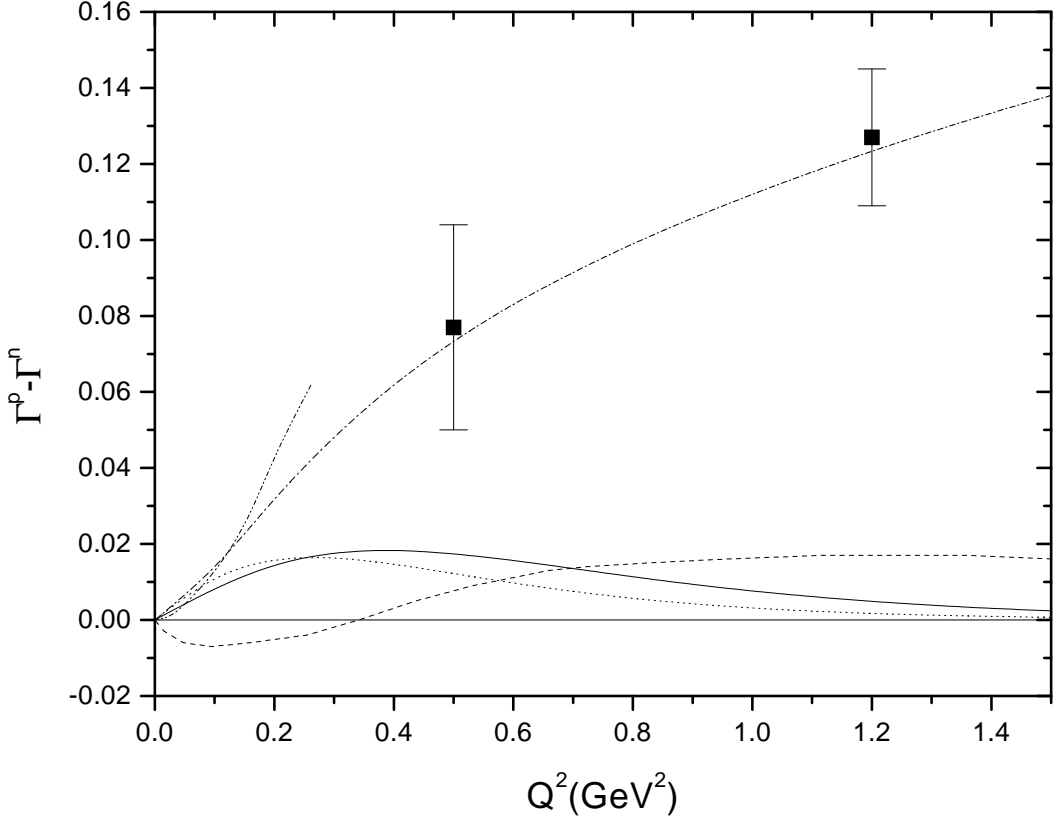, height=13cm} \caption{The Bjorken
integral $\Gamma_p-\Gamma_n$of the spin structure function
$g_1\left(x, Q^2\right)$. The solid and dotted lines show our
calculation with and without two-body exchange currents. The
dashed line is the result of Ref.\protect\cite{burkert93}. The
dashed-double dotted line is the $\chi$PT
prediction\protect\cite{ji00}. Our result with the non-resonant
part and the two-body exchange currents contributions is shown by
the dashed-dotted line. The experimental data are from
\protect\cite{E14398}. }\label{gamma}
\end{center}
\end{figure}

For the proton or neutron separately, the contribution of the
resonance $P_{33}\left(1232\right)$ is the dominant one. However,
for the difference between the proton and the neutron, the
contribution of this resonance is absent. In this case, chiral
perturbation theory is useful, but it is valid only to as low as
$0.2~{\rm GeV}^2$\cite{ji00}. In Fig. \ref{gamma}, we show our
results for $\Gamma^p-\Gamma^n$, and we also compare them with the
results of chiral perturbation theory\cite{ji00}. Our results with
only the resonant part have the correct slope and sign in the low
$Q^2$ region.

\begin{figure}
\begin{center}
\epsfig{figure=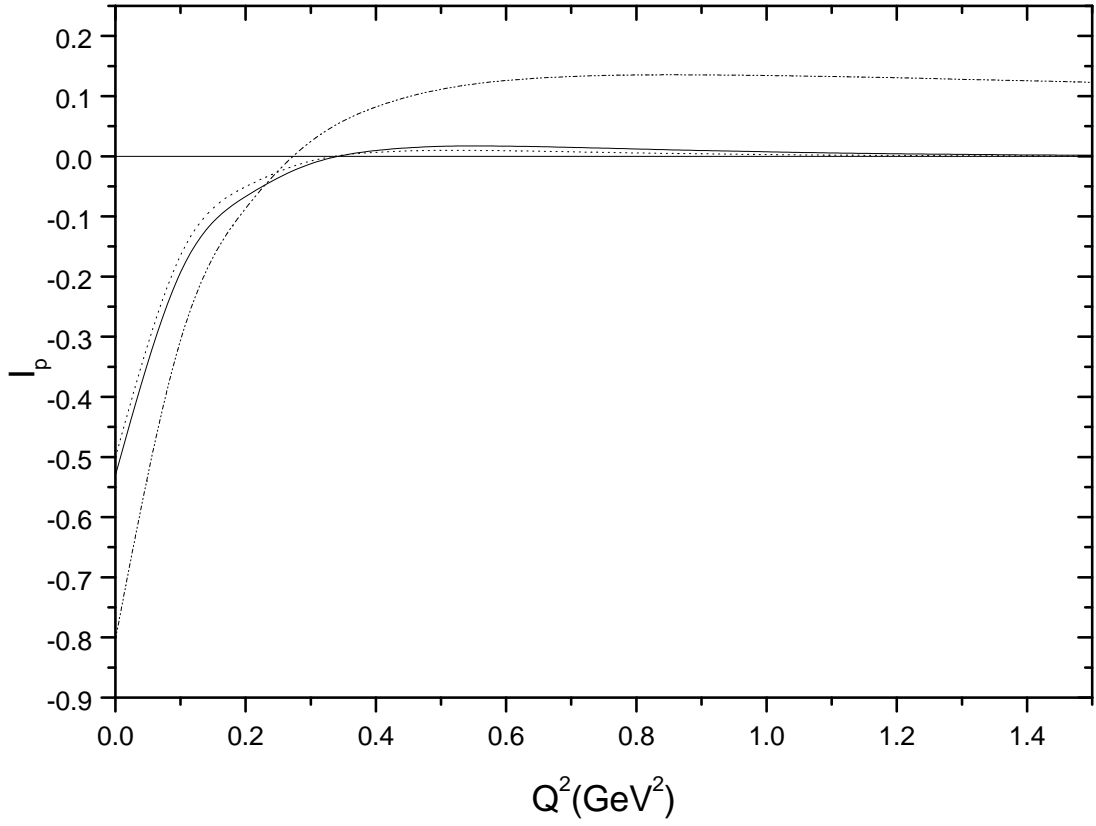, height=13cm} \caption{The $Q^2$
dependence of GDH sum rule for proton. The solid and dotted lines
show our calculation with and without the two-body exchange
currents for proton. Our result with the non-resonant part and the
two-body exchange currents contributions is shown by the
dashed-dotted line.}\label{gdhp}
\end{center}
\end{figure}

\begin{figure}
\begin{center}
\epsfig{figure=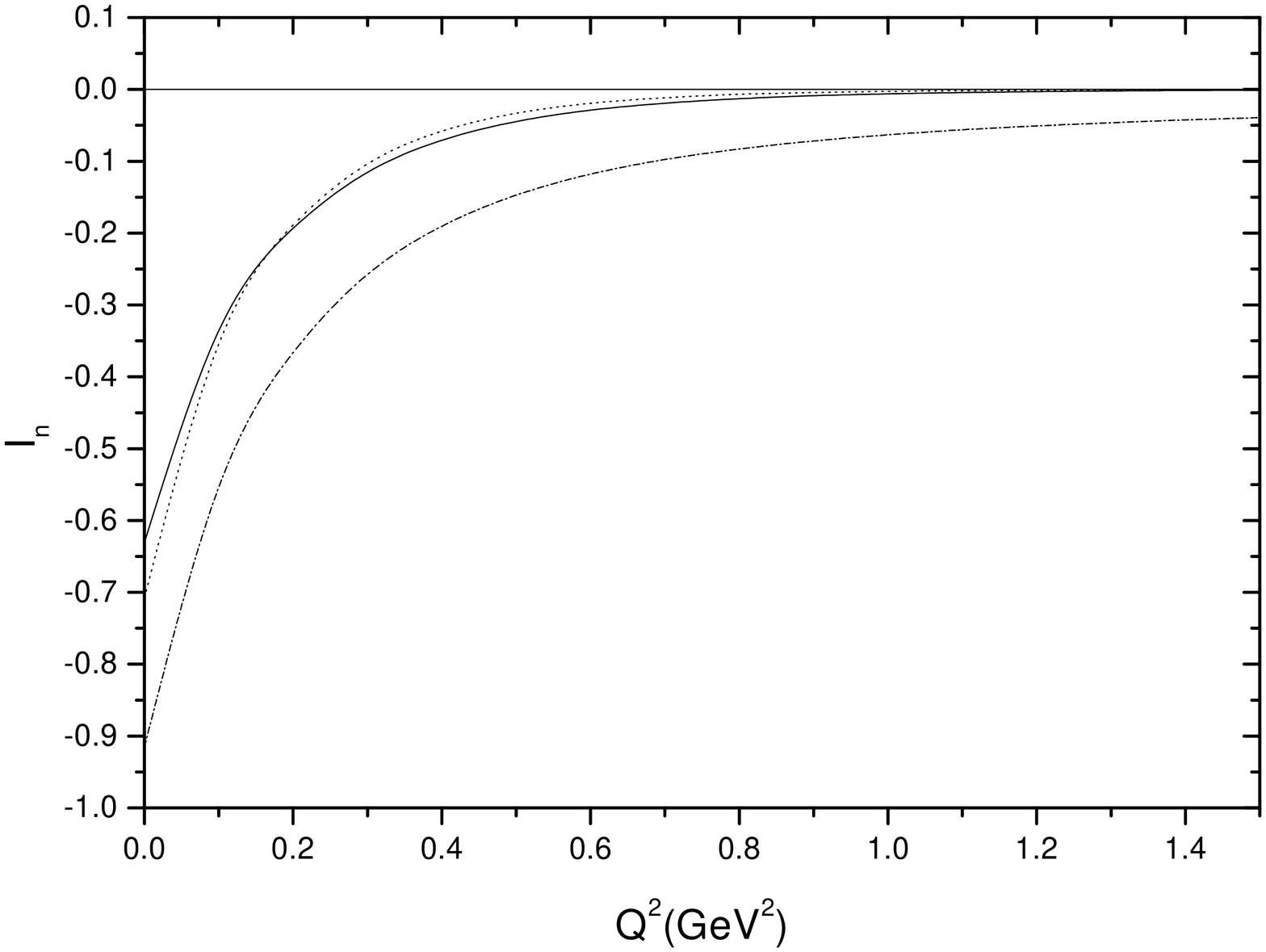, height=13cm}
\caption{ The same as Fig. \ref{gdhp} for neutron. }\label{gdhn}
\end{center}
\end{figure}

In Figs. \ref{gdhp} and \ref{gdhn}, we show our results for the
$Q^2$ dependence of the GDH sum rule, both for the proton and the
neutron, with the following extension:
\begin{equation}
I_{GDH}\left(Q^2\right)=\frac{2M_T^2}{Q^2}\int^{x_0}_0
g_1\left(x,Q^2\right) dx.
\end{equation}
In our calculations, the sign-changing point of the generalized
GDH sum rule for the proton appears around $Q^2\sim 0.27 {\rm
GeV}^2$.

\section{summary}

Based on a chiral constituent quark model, we have calculated the
photo-production amplitudes of all the nucleon resonances with
masses below $2.0~{\rm GeV}$, and the first moment of the nucleon
spin structure function $g_1\left(x, Q^2\right)$, including both
relativistic corrections and those two-body exchange currents
corresponding to the one-gluon exchange, one-pion exchange,
one-sigma exchange and confinement potentials.

We find that the two-body exchange currents required by gauge
invariance, specially the exchange current corresponding to the
confinement potential, play an important role in most cases. But
for $L=2$ nucleon resonances, the two-body exchange currents are
less important. After taking into account the effects of both the
two-body exchange currents and the relativistic corrections, the
fit to experiment is improved.

It is known that the first moment of the nucleon spin structure
function $g_1\left(x, Q^2\right)$ is strongly $Q^2$ depended below
$Q^2=1.0 ~{\rm GeV}^2$, and we obtain a sign change at
$Q^2=0.27~{\rm GeV}^2$. This strong $Q^2$ dependence means that in
order to obtain a reliable theoretical prediction, the saturation
of resonances should be performed carefully, and that is what we
have done in this paper.

We mainly concentrated on the effects of the two-body exchange
currents and relativistic effects, and did not consider
configuration mixing effects. However, it is known that these
effects have some contribution to resonance
photo-production\cite{li90}. Furthermore, in order to compare with
experiment, the $q^2$ dependence of helicity amplitudes also needs
to be calculated. The confinement exchange current is also model
dependent, and it is necessary to consider other possibilities,
such as linear confinement and color screening potentials. All
these effects will be included in future work.

\acknowledgments{
This research is supported by Fondecyt (Chile) postdoctoral fellowship
3000055 and project number 8000017, by a C\'atedra Presidencial (Chile),
and partly by the post Dr. foundation of SED of China.}

\begin{table}
\begin{center}
\caption{Photo-production amplitudes of [$70,1^-$] resonance
states in center mass frame in units of ${\rm GeV}^{-1/2}\times
10^{-3}$. $A_T$ shows the sum of all contributions. The
experimental data are given by the most recent Particle Data
Review\protect\cite{pdg}. }
\begin{tabular}{rrrrrrrrrrrr}\label{tab1}
Multiple States & $A^N_J$ & $A_{NR}$ & $A_{SO}$ & $A_{NA}$ & $A_{gq\bar{q}}$
& $A_{\gamma\pi\pi}$ & $A_{\pi q\bar{q}}$ & $A_\sigma$ & $A_{conf}$ & $A_T$ & Exp.\\
$\left[70,1^-\right]_1\, S_{11}\left(1535\right)$ & $A^p_{1/2}$
& 151 & -43 & 30 & -21 & 15 & -16 & 5 & -41 & 80 & 90$\pm$30 \\
 & $A^n_{1/2}$
& -99 & 14 & 0 & 0 & -15 & 16 & -2 & 14 & -72 & -46$\pm$27 \\
$D_{13}\left(1520\right)$ & $A^p_{3/2}$
& 94 & 27 & -19 & 22 & 14 & -20 & 0 & -1 & 117 & 166$\pm$5 \\
 & $A^n_{3/2}$
& -93 & -9 & 0 & -22 & -14 & 20 & 0 & 0 & -118 & -139$\pm$11 \\
 & $A^p_{1/2}$
& -59 & 15 & -11 & -6 & 8 & -12 & -7 & 59 & -13 & -24$\pm$9 \\
 & $A^n_{1/2}$
& -16 & -5 & 0 & -3 & -8 & 11 & 2 & -20 & -39 & -59$\pm$9 \\
$S_{11}\left(1650\right)$ & $A^p_{1/2}$
& 0 & 0 & 18 & 18 & 13 & -29 & 0 & 0 & 20 & 53$\pm$16 \\
 & $A^n_{1/2}$
& 29 & -12 & -4 & -15 & -17 & 29 & 2 & -14 & -2 & -15$\pm$21 \\
$D_{13}\left(1700\right)$ & $A^p_{3/2}$
& 0 & 0 & -33 & -36 & 13 & -15 & 0 & 0 & -71 & -2$\pm$24 \\
 & $A^n_{3/2}$
& -68 & 22 & 9 & 27 & -3 & 15 & -5 & 34 & 31 & -3$\pm$44 \\
 & $A^p_{1/2}$
& 0 & 0 & -19 & -17 & 0 & -12 & 0 & 0 & -48 & -18$\pm$13 \\
 & $A^n_{1/2}$
& -12 & 12 & 5 & 16 & 2 & 12 & -1 & 6 & 40 & 0$\pm$50 \\
$D_{15}\left(1675\right)$ & $A^p_{3/2}$
& 0 & 0 & 0 &  -7 & 3 & -1 & 0 & 0 & -5 & 15$\pm$9 \\
 & $A^n_{3/2}$
& -57 & 0 & 0 & 0 & 5 & 0 & -4 & 28 & -28 & -58$\pm$13 \\
 & $A^p_{1/2}$
& 0 & 0 & 0 & -5 & 2 & -1 & 0 & 0 & -4 & 19$\pm$8 \\
 & $A^n_{1/2}$
& -41 & 0 & 0 & 0 & 4 & 0 & -3 & 20 & -20 & -43$\pm$12 \\
$S_{31}\left(1650\right)$ & $A^{p,n}_{1/2}$
& 35 & 13 & 8 & 9 & 18 & -17 & -2 & 14 & 78 & 27$\pm$11 \\
$D_{33}\left(1700\right)$ & $A^{p,n}_{3/2}$
& 66 & -7 & -6 & 0 & 13 & -33 & 0 & 0 & 33 & 85$\pm$22 \\
 & $A^{p,n}_{1/2}$
& 81 & -4 & -3 & -11 & 14 & -19 & 3 & -21 & 40 & 104$\pm$15 \\
$\left[70,0^+\right]_2\, P_{11}\left(1710\right)$ & $A^p_{1/2}$
& -25 & 28 & -2 & -6 & 0 & -1 & -4 & -24 & -34 & 9$\pm$22 \\
 & $A^n_{1/2}$
& 8 & -9 & 0 & -1 & 0 & -1 & 1 & 8 & 6 & -2$\pm$14 \\
\end{tabular}
\end{center}
\end{table}

\begin{table}
\begin{center}
\caption{Photo-production amplitudes of [$56,2^+$] resonance
states in center mass frame in units of ${\rm GeV}^{-1/2}\times
10^{-3}$. $A_T$ shows the sum of all contributions. The
experimental data are given by the most recent Particle Data
Review\protect\cite{pdg}. }
\begin{tabular}{rrrrrrrrrrrr}\label{tab2}
Multiple States & $A^N_J$ & $A_{NR}$ & $A_{SO}$ & $A_{NA}$ & $A_{gq\bar{q}}$
& $A_{\pi q\bar{q}}$ & $A_{\gamma\pi\pi}$ & $A_\sigma$ & $A_{conf}$ & $A_T$ & Exp.\\
$\left[56,2^+\right]_2\, P_{13}\left(1720\right)$ & $A^p_{3/2}$
& 32 & -26 & 2 & 5 & -8 & -1 & 0 & 5 & 9 & -19$\pm$20 \\
  & $A^n_{3/2}$
& 0 & 18 & 2 & -2 & 8 & 4 & 0 & -3 & -13 & -29$\pm$61 \\
 & $A^p_{1/2}$
& -116 & 46 & 4 & 12 & 0 & -6 & -2 & 34 & -28 & 18$\pm$30 \\
 & $A^n_{1/2}$
& 41 & -31 & 4 & -4 & 0 & 8 & 2 & -23 & -3 & 1$\pm$15 \\
$F_{15}\left(1680\right)$ & $A^p_{3/2}$
& 64 & 35 & 3 & 9 & 4 & -5 & 0 & -6 & 104 & 133$\pm$12 \\
 & $A^n_{3/2}$
& 0 & -23 & -3 & -3 & -4 & 5 & 0 & 4 & -24 & -33$\pm$9 \\
 & $A^p_{1/2}$
& -28 & 25 &  2 & -1 & 0 & 0 & -3 & 46 & 41 & -15$\pm$6 \\
 & $A^n_{1/2}$
& 49 & -16 & -2 & 0 & 0 & 0 & 2 & -31 & 2 & 29$\pm$10 \\
$P_{31}\left(1910\right)$ & $A^{p,n}_{1/2}$
& -26 & 0 & 0 & -3 & -6 & -4 & -1 & 16 & -24 & 3$\pm$14 \\
$P_{33}\left(1920\right)$ & $A^{p,n}_{3/2}$
& 45 & -20 & 2 & -3 & 4 & 5 & 2 & -25 & 10 & 23$\pm$17 \\
 & $A^{p,n}_{1/2}$
& -26 & 34 & 3 & 0 & 5 & 0 & -1 & 10 & 25 & 40$\pm$14 \\
$F_{35}\left(1905\right)$ & $A^{p,n}_{3/2}$
& -71 & 49 & 4 & 1 & 3 & -3 & -3 & 36 & 16 & -45$\pm$20 \\
 & $A^{p,n}_{1/2}$
& -17 & 34 & 3 & 3 & 4 & -5 & -4 & 4 & 22 & 25$\pm$11 \\
$F_{37}\left(1950\right)$ & $A^{p,n}_{3/2}$
& -55 & 0 & 0 & -3 & -3 & 5 & -3 & 34 & -25 & -97$\pm$10 \\
 & $A^{p,n}_{1/2}$
& -42 & 0 & 0 & -3 & -2 & 4 & -3 & 26 & -20 & -76$\pm$12 \\
$\left[56,0^+\right]_2\, P_{11}\left(1440\right)$ & $A^p_{1/2}$
& 38 & -34 & 7 & -20 & -15 & 5 & -19 & -27 & -65 & -55$\pm$4 \\
 & $A^n_{1/2}$
& -25 & 23 & -7 & 7 & 15 & -9 & 13 & 18 & 35 & 40$\pm$10 \\
$P_{33}\left(1600\right)$ & $A^{p,n}_{3/2}$
& -47 & 36 & -8 & -10 & 11 & -19 & -17 & -9 & -63 & -9$\pm$21 \\
 & $A^{p,n}_{1/2}$
& -27 & 21 & -5 & -6 & 6 & -11 & -10 & -5 & -37 & -23$\pm$20 \\
$\left[56,0^+\right]_0\, P_{33}\left(1232\right)$ & $A^{p,n}_{3/2}$
& -180 & -12 & -12 & -19 & -33 & 27 & -21 & 68 & -182 & -255$\pm$8 \\
 & $A^{p,n}_{1/2}$
& -104 & -7 & -7 & -11 & -19 & 16 & -12 & 39 & -105 & -135$\pm$5 \\
\end{tabular}
\end{center}
\end{table}

\end{document}